\documentclass[a4paper,fleqn,usenatbib]{mnras}
\usepackage[T1]{fontenc}
\usepackage{ae,aecompl}
\usepackage{multirow,ulem}

\usepackage{graphicx}	
\usepackage{amsmath}	
\usepackage{amssymb}	
\usepackage{natbib}
\usepackage{breqn}
\usepackage{amsmath}
\usepackage{siunitx}
\usepackage[dvipsnames]{xcolor}

\newcommand{\vect}[1]{\boldsymbol{#1}}

\PassOptionsToPackage{dvipsnames}{xcolor}
\definecolor{webgreen}{rgb}{0,.5,0}
\newcommand{\ufhref}[3][blue]{\href{#2}{\color{#1}{#3}}}%

\newcommand{\PS}[2]{{\color{magenta}\sout{#1}#2}}

\definecolor{bb}{cmyk}{0,0.5,1,0}
\newcommand{\SG}[1]{{\color{bb} #1}}
\newcommand{\SGC}{\bgroup\markoverwith{\color{bb}{\rule[0.5ex]{2pt}{0.4pt}}}\ULon}

\definecolor{arsenic}{rgb}{0.23, 0.27, 0.29}
 
\def\blue{\textcolor{blue}}

\title[Cosmic rays from massive star clusters : a close look at Westerlund 1]{Cosmic rays from massive star clusters : A close look at Westerlund 1}
\author[Bhadra etc]
{
Sourav Bhadra$^{1,2}$, Siddhartha Gupta$^{3}$, Biman B. Nath$^{1}$, Prateek Sharma$^{2}$ \\
\footnotesize \it $^{1}$Raman Research Institute, Sadashiva Nagar, Bangalore 560080, India; \href{mailto:sbhadra@rri.res.in}{sbhadra@rri.res.in}\\ 
\footnotesize \it $^{2}$Joint Astronomy Programme, Department of Physics, Indian Institute of Science, Bangalore 560012, India;
\href{mailto:souravbhadra@iisc.ac.in}{souravbhadra@iisc.ac.in}\\ 
\footnotesize \it $^{3}$Department of Astronomy \& Astrophysics, University of chicago, IL 60637, USA \\ 
}

\date{Accepted XXX. Received YYY; in original form ZZZ}

\pubyear{2020}

\begin{document}
\label{firstpage}
\pagerange{\pageref{firstpage}--\pageref{lastpage}}
\maketitle


\newcommand{\3}{\ss}
\newcommand{\n}{\noindent}
\newcommand{\eps}{\varepsilon}
\def\be{\begin{equation}}
\def\ee{\end{equation}}
\def\ba{\begin{eqnarray}}
\def\ea{\end{eqnarray}}
\def\de{\partial}
\def\msun{M_\odot}
\def\div{\nabla\cdot}
\def\grad{\nabla}
\def\rot{\nabla\times}
\def\ltsima{$\; \buildrel < \over \sim \;$}
\def\simlt{\lower.5ex\hbox{\ltsima}}
\def\gtsima{$\; \buildrel > \over \sim \;$}
\def\simgt{\lower.5ex\hbox{\gtsima}}
\def\etal{{et al.\ }}





\begin{abstract}
\setlength{\leftskip}{10em}
We study the effect of cosmic ray (CR) acceleration in the massive compact star cluster Westerlund 1 in light of its recent detection in $\gamma$-rays. 
Recent observations reveal a $1/r$ radial distribution of the 
CR energy density. Here we theoretically investigate whether or not this profile can help to distinguish between (1) continuous CR acceleration in the star cluster stellar wind-driven shocks and (2) discrete CR acceleration in multiple supernovae shocks -- which are often debated in the literature.
Using idealized two-fluid simulations and exploring different acceleration sites and diffusion coefficients, we 
obtain the CR energy density profile and luminosity to find the best match for the $\gamma$-ray observations. We find that the inferred CR energy density profiles from observations of $\gamma$-ray luminosity and mass can be much different from the true radial profile. CR acceleration at either the cluster core region or the wind termination shock can explain the observations, if the diffusion coefficient is $\kappa_{\rm cr}\sim  10^{27}$ cm$^2$ s$^{-1}$ and a fraction of $\approx 10\%-20\%$ of the shock power/post-shock pressure is deposited into the CR component. 
We also study the possibility of discrete supernovae (SN) explosions being responsible for CR acceleration and find that with an injection rate of 1 SN in every $\sim 0.03$ Myr, one can explain the observed $\gamma$-ray profile. This multiple SN scenario is consistent with X-ray observations only if the thermal conductivity is close to the Spitzer value.
\end{abstract}

\begin{keywords}
\setlength{\leftskip}{10em}
ISM: Cosmic rays, bubbles---gamma-rays: diffuse background, ISM
\end{keywords}

\section{Introduction}
There is a recent surge of interest in the acceleration of cosmic rays (CRs) in massive star clusters, which are increasingly seen as a supplementary site of CR acceleration in our Galaxy besides isolated supernova remnants.
The $\gamma$-ray observations by {\it Fermi-LAT, HESS,} and {\it HAWC} have provided the evidence of hadronic acceleration in a handful of massive star clusters. For instance, the Cygnus OB association, Westerlund 1, Westerlund 2 in our Galaxy and 30 Doradus in the Large Magellanic Cloud are some of the bright sources of $\gamma$-rays from GeV to several TeV energies and have been interpreted as powerful CR accelerators (\citealt{Abdo2010,Ackermann2011,Abramowski2015,Abeysekara2021}).
Although star forming regions have been previously discussed as possible sources of CRs  (e.g., \citealt{Knodlseder2013,Bykov2014, Aharonian2018}), these $\gamma$-ray observations  have strengthened this  hypothesis and they allow us to make more detailed theoretical models, thereby to improve our understanding of CR acceleration in these environments.

The idea that massive star clusters are potential acceleration sites of CRs has long been discussed in the literature, beginning with energetic arguments \citep{cesarsky1983}.
CR acceleration in these environments can also solve many problems associated with the isolated supernova paradigm, including proton acceleration up to $\sim$ PeV energies (representing the knee of the Galatic CR spectrum) and the excess of $^{22}$Ne/$^{20}$Ne in CRs compared to the standard ISM composition (\citealt{Higdon2003}, for a review, see \citealt{Gabici2019}). Detailed theoretical investigations support these predictions (e.g., \citealt{Gupta2020}, \citealt{Morlino2021}).
In this regard, \cite{Gupta2020} have demonstrated that the problem of the large observed ratio of Neon isotopes ($^{22}$Ne/$^{20}$Ne) can be solved by invoking CR acceleration in the stellar winds in star clusters.
Concurrently, it has been suggested from various phenomenological considerations that most of the observed CR grammage in the Galaxy is accumulated in star clusters, and not while propagating through the interstellar medium (ISM) at large (see, e.g., \citealt{blasi2009, cowsik2016, eichler2017, biermann2018}). Taking this cue, \cite{Nath2020} have shown that the resulting $\gamma$-rays from star clusters can explain a significant fraction of the observed diffuse Galactic background.
These developments prod us to look deeper at the individual and detailed observations of star clusters. 

Recently, using the $\gamma$-ray observations of Cygnus and Westerlund 1 (hereafter referred to as Wd1) and CO/HI observations, \blue{Aharonian et al. 2019} (hereafter, \citealt{Aharonian2019}) reported that the spatial distribution of CR energy density in these objects follows $1/r$  profile. 
They suggested a steady injection {(over $\sim $few Myr)} of CRs 
instead of instantaneous injection as normally expected in case of an isolated supernova. 
 Although such profiles can be derived by solving steady-state CR transport equation (cf. \S \ref{sec:preli}), it is worth mentioning that the steady-state assumption is questionable when the shock-bubble structure continuously evolves. We subject these observations to scrutiny with two-fluid hydro simulation and check if other interpretations (of the actual CR energy density profile and the mode of CR injection) 
are ruled out and if the observations can be used to infer the relevant physical parameters for CR acceleration. 
In the present work, we have studied different CR injection methods in order to understand the observed $\gamma$ -ray luminosity, mass and CR energy density of the Wd1 cluster. We selected Wd1 for our study mainly because it is a compact cluster and can be modeled convincingly using 1-D simulations. Although there are a few other clusters that have been detected in $\gamma$-rays (such as Cygnus \citealt{Bartoli2014}), those objects are distinctly non-spherical in morphology and have substructures, which make them difficult to compare with 1-D simulations. We begin with analytical estimates of $\gamma$-ray luminosity in section 2. The numerical simulation setup is described in section 3. In section 4, we present our results, followed by further discussions in section 5, and we summarize in section 6. 


\section{Preliminaries}\label{sec:preli}
In a couple of massive star clusters (e.g., Wd1 cluster; see \citealt{Aharonian2019}), the CR energy density in different annuli, as estimated from $\gamma$-ray luminosity, 
has been found to follow a $1/r$ profile.
These profiles are often interpreted  
in terms of a steady injection of CRs (with energy $E$) from the dense core of compact star cluster with an energy-dependent CR diffusion (diffusion coefficient $\kappa_{\rm cr} (E)$). 
 This can be shown directly by using the CR diffusion-transport equation: 
\be
\frac{\partial N (E)}{\partial t}= \kappa_{\rm cr}(E) \nabla ^2 N(E) + Q(E),
\ee
where, $N(E)$ is the number density of CRs and $Q(E)$ is their energy injection rate density. The energy moment of this equation can be written as 
\be
{\partial e_{\rm cr} \over \partial t}=\kappa_{\rm cr} \nabla^2 e_{\rm cr} + {L \over V},
\ee
where $e_{\rm cr}$ represents CR energy density, $\kappa_{\rm cr}$ is an appropriately averaged diffusion coefficient for CRs, and $\frac{L}{V}$ the CR luminosity density. Let us assume that CR particles are injected in a small central region of radius $r_0$, which is much
smaller in extent than the size of the star cluster. The rest
of the volume is assumed to be devoid of CR production sites for simplicity. Therefore, except in the very central region, we need to solve the equation
\be
\frac{\partial e_{\rm cr}} {\partial t} = \kappa_{\rm cr} \nabla ^2 e_{\rm cr} \,.
\label{trans1}
\ee
In steady state, it reduces to (in spherical symmetry)
\be
\frac{d}{dr} \Bigl ( r^2 \frac{d}{dr} e_{\rm cr} \Bigr ) =0\,, \, \Rightarrow e_{\rm cr} \propto \int \frac {dr}{r^2}\,,
\ee
which has the solution $e_{\rm cr} = \frac{A}{r} +B$, where $B\rightarrow 0$ since the CR energy density is zero at infinity,
and where $A$ is a constant that depends on the boundary condition at $r_0$. This is the $1/r$ solution which is taken to be an evidence for steady injection of CR energy in massive clusters \citep{Aharonian2019}.

Clearly, the above estimate neglects some crucial aspects such as (1) advection of CRs, (2) role of CR acceleration sites other than the central region, (3) losses due to 
radiative cooling of the gas, (4) projection effects, and (5) dominating $\gamma$-ray emission regions (which can be different from acceleration site; cf. \ref{diff_injection}). These considerations are important for the case of massive star clusters. 
Therefore, time-dependent numerical simulations are essential. 
Our two-fluid approach allows us to consider the wind termination shock as an acceleration site, as well as to study the effect of time-varying CR/mechanical luminosity of the star cluster, including discrete SNe. 

\subsection{$\gamma$-ray luminosity ($L_{\gamma}$)} \label{subsec:gam-lum}
\subsubsection{Hadronic contribution}
One of the major sources of $\gamma$-rays in Wd1 is the hadronic interaction between CR protons and protons in the ambient gas (see below for an estimate of $\gamma$-ray flux in the leptonic case, from inverse Compton scattering of stellar photons by CR electrons). The mechanism of production of $\gamma$-rays is
\begin{equation}
  p+p  \rightarrow  p+p+\pi^0\,, \quad
  \pi^0  \rightarrow  \gamma + \gamma. 
\end{equation}
Therefore, observations of $\gamma$ ray photons hold clue to the spatial distribution of cosmic ray protons. 

To estimate $\gamma$-ray luminosity due to hadronic interactions, we use the prescription of Dermer’s model (\citealt{dermer86,pfrommer04}), which yields the luminosity between $(E_{\gamma1}$ and $E_{\gamma2})$ energies: 
\begin{eqnarray}
    L_\gamma ^H &=& \int_V dV \,  \int_{E_{\gamma1}}^{E_{\gamma2}} dE_{\gamma} \, E_{\gamma} \, q_{\gamma}(n_N, e_{\rm cr}, E_{\gamma}) \nonumber \\
        &=& \int_V dV \, n_N(r) \, e_{\rm cr}(r) \left [ \int_{E_{\gamma1}}^{E_{\gamma2}} dE_{\gamma} \, E_{\gamma} \, \tilde{q_{\gamma}} \right].
\label{eq:sourcefn}
\end{eqnarray}
Here, $q_{\gamma} = dN/(dt\,dV\,dE_{\gamma})$ is the number of $\gamma$-ray photons emitted per unit volume per unit time per unit energy, which is proportional to the number density of target nucleon ($n_N$) and the CR energy density ($e_{\rm cr}$). The integration is to be carried over the entire volume of the emission region. The isotropic source function $\tilde{q_{\gamma}}$, used in the second integral, is given as (see e.g., \citealt{pfrommer04,Gupta2018,jana20}):
\begin{equation}
    \tilde{q_{\gamma}}= \frac{\sigma_{pp} c \left(\frac{E_{\pi^0}}{GeV}\right)^{-\alpha_{\gamma}}\left[\left(\frac{2E_{\gamma}}{E_{\pi^0}}\right)^{\delta_{\gamma}}+\left(\frac{2E_{\gamma}}{E_{\pi^0}}\right)^{-\delta_{\gamma}}\right]^{-\frac{\alpha_{\gamma}}{\delta_{\gamma}}}}     {\xi^{\alpha_{\gamma}-2}\left(\frac{3\alpha_{\gamma}}{4}\right) \left(\frac{E_p}{(2\alpha_p-2) {\rm GeV}}\right)\left(\frac{E_p}{\rm GeV}\right)^{1-\alpha_p}\beta\left(\frac{\alpha_p-2}{2}, \frac{3-\alpha_p}{2}\right)}.
\end{equation}
Here, $\xi = 2$ 
is the multiplicity factor, which denotes two leading pion jets leaving the interaction site, $E_p$ and $E_{\pi^0}$ are the rest mass energy of proton and pions ($\pi^0$) respectively. The spectral indices of 
the incident CR protons and emitted $\gamma$-ray photons are denoted by
$\alpha_p$ and $\alpha_{\gamma}$ respectively, $\delta_{\gamma}= 0.14\alpha_{\gamma}^{-1.6}+0.44$ is the spectral shape parameter and $ \sigma_{pp}=32(0.96+ e^{4.4-2.4\alpha_{\gamma}})$
mbarn (for details see equations (8) and (19)–(21) in \citet{pfrommer04}). We use $\alpha_{\gamma}=\alpha_p=2.3$ following the spectral fit of \citet{Ackerman15}. The integration 
over the $\gamma$-ray photon energy in equation \ref{eq:sourcefn} for $E_{\gamma1}=$ 1 TeV and $E_{\gamma2}=100$ TeV (the contribution at higher energies is very small)
gives $1.05 \times 10^{-17}$ \si{cm^3 s^{-1}}. Thus the $\gamma$ -ray luminosity above $1$ \si{TeV} can be written as,
\begin{equation}
    L_{\gamma} ^H \sim 10^{-17} \left(\frac{\Delta V}{\si{cm^3}}\right) \left(\frac{n_N}{\si{cm^{-3}}}\right) \left(\frac{e_{\rm cr}}{\si{erg \, cm^{-3}}}\right) \si {erg \, s^{-1}}\,.
    \label{eq:lgamma}
\end{equation}

We use this equation to calculate the $\gamma$-ray luminosity from the relevant region of the cluster. On inverting equation \ref{eq:lgamma} we get the CR energy density above $10$ TeV, 
\begin{eqnarray}
     e_{\rm cr} (> 10 {\rm TeV}) \approx 1.5 \times 10^{-2} \left( \frac{L_{\gamma}^H}{10^{34} \, \si{erg \, s^{-1}}} \right) \left(\frac{10^6 M_{\odot}}{M} \right)\, \nonumber\\
     \si{eV \, cm^{-3}} \, .
     \label{eq:CR_10TeV}
\end{eqnarray}
where $M$ is the mass and $L_{\gamma}$ is the $\gamma$-ray luminosity above $1$ TeV energy. 

\subsubsection{Leptonic contribution}
 It is also possible to have a leptonic contribution to the total $\gamma$-ray luminosity, from inverse Compton scattering of stellar radiation photons by CR electrons, especially 
 close to the star cluster where stellar photon density is significant. 
 We estimate the leptonic emission as follows.
 
 The energy density of CR electrons is assumed to be $0.01$ of the total CR energy density (i.e $e_{\rm cr,e} \approx 0.01 e_{\rm cr}$). This value has some uncertainty. From observations in the solar system, at CR energy $\sim 10$ \si{GeV}, where solar modulation effects are low, the ratio of CR electron to proton energy is known to be $1\%$ (\citet{Longair2011} section 15.1, \citet{Schlickeiser2002}). Assuming the energy distribution of CR electrons to be $n(\Gamma)=\kappa_1 \Gamma^{-p}$ (in terms of the Lorentz factor $\Gamma$), where $p=2.3$ (same as that of protons), the normalization constant $\kappa_1$ is given by
\begin{equation}
\kappa_1\approx {e_{\rm cr,e} \over m_e c^2} (p-2) \Bigl [ {1 \over \Gamma_L ^{p-2}} - {1 \over \Gamma_U ^{p-2}} \Bigr ] ^{-1}\,.
 \label{eq:kappa}
\end{equation}
Here, the upper cutoff to the Lorentz factor can be taken as $\Gamma_U \rightarrow\infty$ and the lower cutoff ($\Gamma_L$), as unity. Then, the total IC luminosity (which provides an upper limit to $\gamma$-ray luminosity) is given by (\citet{rybicki}, eqn 7.21),
\begin{equation}
    L_\gamma ^{IC} =\int_V \, dV \, \Bigl [ {4 \over 3} \sigma_T \, c \, e_{\rm ph} \, \kappa_1 \, {\Gamma_{\rm max}^{3-p} -\Gamma_{\rm min} ^{3-p} \over 3-p} \Bigr ]\,,
     \label{eq:linverse}
\end{equation}
where, $e_{\rm ph}$ is the photon energy density, which at a distance $r$ from the central core region of star cluster is given by
\begin{equation}
    e_{\rm ph} = \frac{L_{\rm rad}}{4 \pi r^2 c}.
    \label{eq:eph}
\end{equation}
Hence one can obtain an upper limit to the leptonic contribution by using equations \ref{eq:kappa}, \ref{eq:linverse} \& \ref{eq:eph}. 

Using these equations, one gets a sharply declining profile of $L_\gamma$ with distance, because of the rapid decline of $e_{\rm ph}$ with radius. This is in contrast with the observed increasing profile of $L_\gamma$ with projected distance. The observed profile, therefore, works against the leptonic interpretation of the origin of $\gamma$-rays. 


Since IC scattering boosts the seed photon energy by a factor of $\Gamma^2$ ($\Gamma$ being the electron Lorentz factor), a seed (stellar) photon of $\sim 1$ eV will require $\Gamma=10^6$ for it to be scattered into $1$ TeV energy. If we take the photons in the waveband $0.01\hbox{--}100$ eV (FIR to FUV), then the total radiation luminosity of the cluster is given by $L_{\rm rad} \sim 500 L_w$ \citep{Leitherer99}, where $L_w$ denotes the mechanical luminosity. In the innermost region considered here, within $9$ pc, the photon energy density amounts to $\approx 1125$ eV/cc. Therefore, electrons that do not cool within $4.5$ Myr have $\Gamma \le 120$, which require a seed photon energy of $\ge 70$ MeV in order to up-scatter to $1$ TeV. Note that, this incident photon energy is much greater than our assumed seed photon energy ($0.01\hbox{--}100$ eV). Therefore no photons in this region can be upscattered to above 1 TeV. 
If we put 
$\Gamma=120$ as $\Gamma_{\rm max}$ and $\Gamma_{\rm min}=1$ in equation \ref{eq:linverse}, we get
\begin{eqnarray}
   L_\gamma ^{IC} \sim  10^{-18} \left(\frac{L_w}{10^{39} \rm erg \, s^{-1}}\right) \times \left[ \int dV \left(\frac{r}{10 \rm pc}\right)^{-2} e_{\rm cr}(r) \right] \nonumber\\
    \rm erg \, s^{-1} \, ,
\end{eqnarray}
where $dV$ and $e_{\rm cr}$ are in cgs unit. A comparison with equation \ref{eq:lgamma} shows that even {\it total} IC losses (only a negligible fraction of this is emitted above 1 TeV) are smaller than the hadronic luminosity above 1 TeV.

\subsection{CR energy density ($e_{\rm cr}$)}
Although our simulation can track the CR energy density ($e_{\rm) cr}$, observations can only determine it through projection, and that too indirectly using $L_\gamma$ and the total projected mass in different projected annuli. In order to compare our calculations with observed parameters, we note that \cite{Aharonian2019} have estimated the CR energy density {$e_{\rm cr, inf}$} above $10$ TeV using the following expression (their equation 7, which is almost identical to our equation \ref{eq:CR_10TeV}),
\begin{equation}
\begin{aligned}
    e_{\rm cr, inf} (>\rm 10TeV) = &1.8 \times 10^{-2} \left(\frac{\eta}{1.5}\right) \left(\frac{L_{\gamma}}{10^{34} \si{erg/s}}\right) \left(\frac{10^6 M_{\odot}}{M}\right) \, \\
    & \hspace{5cm}\si{eV \, cm^{-3}},
    \label{eq:ecr_inf}
\end{aligned}
\end{equation}
where $M$ is the mass and $L_{\gamma}$ is the $\gamma$-ray luminosity above $1$ TeV energy. We use the subscript 'inf' to emphasize that this is the {\it inferred} value of CR energy density, in order to distinguish from the real value, which we get from simulation. 
$\eta$ accounts for nuclei heavier than hydrogen in both cosmic rays and ISM. Clearly, the value of $\eta$ depends on the chemical composition of the ambient gas and CRs. 
The composition parameter $\eta$ varies between $1.5$ to $2$ (\citet{kafexhiu14}, \citet{der1986}), and here we have used $\eta=1.5$. Note that, we mainly consider those CRs which have energy more than $10$ \si{TeV} in our calculations. Also note that the equations \ref{eq:CR_10TeV} and \ref{eq:ecr_inf} are in good agreement. 


\subsection{Distance to Wd1: recent updates and estimation of age}
\label{sect:distance}
There has been an uncertainty regarding the distance to the Wd1 cluster. \cite{Aharonian2019} have used a distance of $4$ kpc. 
However, the recent Gaia Early Data Release 3 (hereafter 'EDR3') \citep{Gaia2021} has provided a more accurate determination of the distance of Wd1, of $2.8$ kpc, which is smaller than previously thought. 
All the distances we use in our simulation, as well as calculations, are based on the  Gaia EDR3 \citep{Gaia2021}. The observed value of projected $\gamma$-ray luminosity, as well as projected mass, also have been modified accordingly. In other words, the physical sizes of the bins have been decreased by a factor of $4/2.8=1.42$.

As far as the age is concerned,  \citet{Gaia2021} stated that the turnoff mass will be reduced from $40$  M$_\odot$ to $22$ M$_\odot$, which would imply an increase in the age. However,  \citet{Negueruela2010}  found the turn off mass to be  $\sim 25$ M$_\odot$, and the age, $4-5$ \si{Myr}. Also, one can estimate the age from the relative number of Wolf-Rayet to Red Supergiants irrespective of the distance, and this yields an age of $4.5-5$ \si{Myr}. Moreover, the age of WD1 cluster cannot be more than $\sim 5$ Myr, 
since Wolf-Rayet stars cannot last longer than this (although \citet{Beasor2021} has claimed a much larger age of $7.2$ Myr). Here, we use an age of $4.5$ Myr, and we show our results at this epoch. The physical and simulation parameters for Wd1 are in table \ref{table:parameters}.

\section{Numerical set up}\label{sec:setup}
We use the publicly available magneto-hydrodynamics code, PLUTO \citep{Mignone2007}, our version of which supports CRs as a fluid detailed in \citealt{Gupta2021}. PLUTO is a finite-volume Godunov code based on Riemann solvers, designed to integrate a system of conservation laws of fluid dynamics that adopts a structured mesh. 
In this work, the code solves the following set of equations:
\begin{subequations}
\begin{eqnarray}
\label{eq:rho}
& \frac{\partial \rho}{\partial t} + \nabla \cdot (\rho \vect{v}) = S_{\rho}, \\ [10pt]
\label{eq:rhov}
& \frac{\partial (\rho \vect{v})}{\partial t} + \nabla \cdot (\rho \vect{v}\otimes \vect{v}) + \nabla (p_{\rm th}+p_{\rm cr})= \rho \vect{g}, \\[10pt]
\nonumber
&\frac{\partial (e_{\rm th}+e_k)}{\partial t} + \nabla [(e_{\rm th}+e_{k})\vect{v}] +\nabla [\vect{v}(p_{\rm th}+p_{\rm cr})] \\
\label{eq:pth}
& = p_{\rm cr} \nabla \cdot \vect{v}- \nabla \cdot \vect{F}_{\rm tc} +		q_{\rm eff}+S_{\rm th} + \rho \vect{v} \cdot \vect{g}, \\[10pt]
\label{eq:ecr}
& \frac{\partial e_{\rm cr}}{\partial t} + \nabla \cdot [e_{\rm cr} \vect{v}] = -p_{\rm cr} \nabla \cdot \vect{v} - \nabla \cdot \vect{F}_{\rm {crdiff}} + S_{\rm cr},
\end{eqnarray}
\end{subequations}
%

where $\rho$ is the mass density, $\vect{v}$ is the fluid velocity, $p_{\rm th}$ and $p_{\rm cr}$ are thermal pressure and CR pressure respectively. $e_{\rm k}$ is the kinetic energy density, $e_{\rm th}= p_{\rm th}/(\gamma_{\rm th}-1)$ and $e_{\rm cr}= p_{\rm cr}/(\gamma_{\rm cr}-1)$ are the thermal energy density and CR energy density, respectively.  $S_{\rho}$, $S_{\rm th}$ and $S_{\rm cr}$ are the mass and energy source terms 
per unit time per unit volume. $\vect{F}_{\rm {tc}}$, $\vect{F}_{\rm {crdiff}}$ represents thermal conduction flux and CR diffusion flux, respectively, $g$ denotes the gravity and $q_{\rm eff}$ accounts for the radiative energy loss of the thermal gas.  We have used HLL Riemann solver, piecewise linear reconstruction and RK2 time stepping. In our simulation, we use a CFL number of $0.4$ and 1-D spherical geometry.\\

\subsection{Ambient medium}
\label{ambient}

In the subsection \ref{subsec:gam-lum} we show that a major fraction of $\gamma$-rays can be produced due to hadronic interactions, and therefore, modelling the gas density of the cloud is crucial.
However, the gas density in these environments is largely uncertain. 
Current observations provide us with the total mass up to a given radius and the projected density profile when the bubble has already evolved. 
With this limited information, we have explored various density distributions and finally selected a density profile (as briefly discussed below), which not only shows a good match with the total gas mass of WD1 \citep{Aharonian2019}, but also gives a size of the bubble at $\sim 4.5$ Myr comparable to observations. 




We use 
a combination of self-gravitating isothermal clouds with solar metallicity following section 4.1 in \citet{Gupta2018}. This gives the total mass density at the central region of the cloud $\sim 625\,m_{\rm H}$ cm$^{-3}$, which drops radially as $\sim 220 (5{\rm pc}/r)\,m_{\rm H}$ cm$^{-3}$; see e.g., their figure 1, giving the mass $\sim 10^6$ $M_{\rm \odot}$ for a cloud of radius $\approx 100$ pc.
These numbers are consistent with WD1. 
Note that, as soon as the wind/SNe becomes active, this initial density profile only remains valid outside the bubble. 
The interior structure evolves depending on the mechanical energy and mass injections from the star cluster, as we discuss in the following sections.

%
%

\begin{table*}
\begin{center}
\renewcommand{\arraystretch}{1.5}
\begin{tabular}{|c| c| c| c|}
    \hline
    \multicolumn{4}{|c|}{\textbf{Westerlund1}}\\
    \hline \hline
    \multicolumn{2}{|c|}{\textbf{Observations}}  & \multicolumn{2}{|c|}{\textbf{Simulation parameters}}\\
    \hline
    \textbf{Extension (pc)} & $60$ & \textbf{$\epsilon_{\rm cr}/ w_{\rm cr}$ range covered}  & $0.1-0.3$\\
    \hline
     \textbf{Age of cluster (Myr)} & $4-6$ & \textbf{$\kappa_{\rm cr}$ range covered} & $(5-100) \times 10^{26}$ cm$^2$ s$^{-1}$\\
    \hline
    \textbf{Kinetic Luminosity $\mathbf{L_w}$ (erg s$^{-1}$)} &  $10^{39}$ & \textbf{Simulation box size} & $250$ pc\\
    \hline
    \textbf{Distance (kpc)}    &  $2.8$ & \textbf{No of grids} & $5000$\\
    \hline
    \textbf{Mass loss rate $\mathbf{\dot{M}}$ ($\mathbf{M_{\odot}/yr}$)}    &  $7.5 \times 10^{-4}$ & \textbf{Cooling} & Tabulated\\
    \hline
\end{tabular}
\end{center}
\caption{\label{tab:table 1} Various physical and simulation parameters of Westerlund 1 used in this work.} 
\label{table:parameters}
\end{table*}

\subsection{Wind driving region}
The main driving engines in star clusters are stellar wind and supernova explosions. While the stellar wind from individual stars can vary with time, the total wind power and mass can be assumed to be constant over time (\citet{Leitherer99}), which are mainly injected by massive stars located in the central regions of compact clusters such as  WD1.
Mass and energy are deposited in a spherical region of radius $r_{\rm inj} = 1$ pc around the centre (the volume of the injection region is $V_{\rm src}=4/3\,\pi r_{\rm inj}^3$) and the spatial
resolution of the runs is $\Delta r = 0.05$ pc. We set this resolution to minimize un-physical cooling losses
(see section 4 in \citet{Sharma14}). The injection region is chosen in such a way that the radiative energy loss
rate is less than the energy injection rate \citep{Sharma14}. In our simulations, we set the mass loss rate
$\dot{M} = 7.5 \times 10^{-4}$ M$_\odot$ yr$^{-1}$ (Table 1) and kinetic luminosity $L_w =
10^{39}$ erg s$^{-1}$. The mass loss rate $\dot{M}$ in chosen so that the wind velocity $v \sim [2L_w/\dot{M}
]^{1/2}$ for Westerlund1 is nearly $2000$ km  s$^{-1}$ \citep{Chevalier85}. An injection parameter $\epsilon_{\rm cr}$
(see equation \ref{eq:epsilon}) is used to specify the fraction of total injected energy given to CRs. The source term
$S_{\rm cr}$ in equation \ref{eq:ecr} can be expressed in terms of the kinetic luminosity of the source region,
\begin{eqnarray}
 S_{\rm cr}= \frac{\epsilon_{\rm cr}L_w}{V_{\rm src}} \,.
\end{eqnarray}
Similarly $S_{\rho}$ (in eqn \ref{eq:rho}) and $S_{\rm th}$ (in eqn \ref{eq:pth}) can be expressed as,
\begin{eqnarray}
 S_{\rho}= \frac{\dot{M}}{V_{\rm src}}\hspace{8pt} ; \hspace{8pt} S_{\rm th}= \frac{(1-\epsilon_{\rm cr})L_w}{V_{\rm src}}\, .
\end{eqnarray}

\subsection{CR injection}
We consider three different methods of CR injection in this paper. In the first case, CRs are injected in the wind driving region, i.e., within $r_{\rm inj}$. In the second case, CRs are injected into the shocked zones. The last case is a combination of both. These injection regions can be seen as possible CR acceleration sites in this object, where the central injection represents unresolved regions, e.g., colliding winds, which can also accelerate CRs \citep{Eichler1993, Bykov2014}. We use the following three conditions to identify the shocked zones (\citealt{Pfrommer2017}, \citealt{Gupta2021}),
\begin{eqnarray}
  \nabla \cdot \vect{v}  &<& 0, \\
  \nabla p \cdot \frac{\Delta r}{p} &>& \delta_{\rm threshold}, \\
  \nabla{T} \cdot \nabla{\rho} &>& 0 \,.
\end{eqnarray}
Here, $\vec{v},~p,~\rho,$ and $T$ are the velocity, pressure, density and temperature of the fluid, respectively. The first condition selects compressed zones, the second condition sets the pressure jump at the shock, and the third condition avoids the contact discontinuity. In the third injection method, we use the combined injection of CRs in the wind driving region as well as in the shocked region. 

For all these different injection methods, the injection of CRs does not add any additional energy in the computational domain. The injection parameter ($\epsilon_{\rm cr}$ or $w_{\rm cr}$, see section \ref{diff_injection}) just distributes a fraction of the total mechanical energy in the CRs either in the wind driving region or in the shocked regions.

\subsection{Microphysics}

\subsubsection{CR Diffusion}
Our simulations include the effects of CR diffusion. For numerical stability, diffusion typically has a much smaller time step than the CFL time step. To make our runs faster, we choose super time-stepping method \citep{Alexiades1996} for the diffusion module, which sub-cycles CR diffusion for each hydro time-step. The CR diffusion flux term can be expressed in terms of CR energy density,
\begin{eqnarray}
\vect{F}_{\rm crdiff} = -\kappa_{\rm cr} \vec{\nabla} e_{\rm cr} \,,
\end{eqnarray}
where $\kappa_{\rm cr}$ is the diffusion coefficient and $e_{\rm cr}$ is the CR energy density. 
Generally, $\kappa_{\rm cr}$ is a function of CR energy, but here we consider a constant value of $\kappa_{\rm cr}$, which can be thought of as its appropriately energy weighted value across the energy distribution function of CRs 
(equation 7 in \citet{Drury1981}).

We use a smaller value for the diffusion coefficient ($\kappa_{\rm cr}$) 
than generally used for the Galactic scales. We set $\kappa_{\rm cr}$ in the range of $(5\hbox{--}100) \times 10^{26}$ cm$^2$
s$^{-1}$. 
This is justified because cosmic rays escaping the acceleration sites are expected to drive turbulence locally, making them diffuse more slowly compared to the ISM at large \citep{Abeysekara2021}. 

\begin{figure*}
\includegraphics[width=\textwidth]{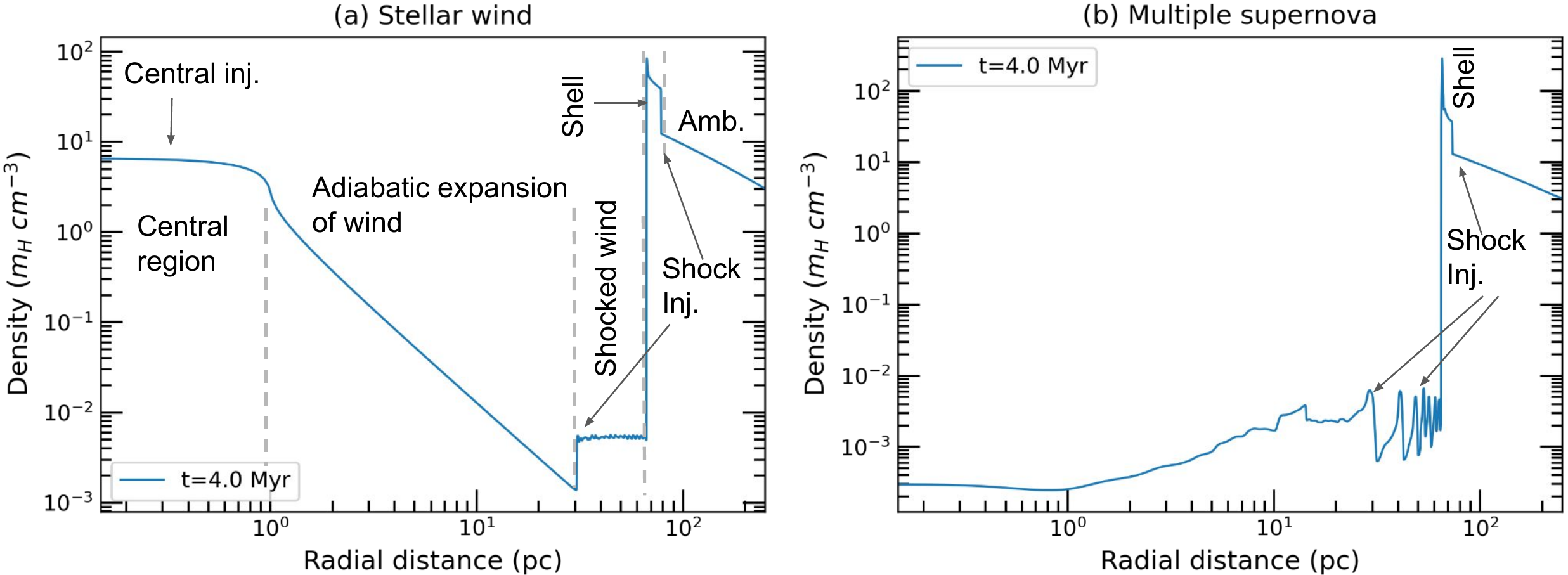}
\caption{ (a) Density profile of a wind-driven bubble 
at a time $t=4.0$ Myr. The horizontal axis represents the distance from the centre in pc and the y-axis denotes the density in terms of \si{m_H \, cm^{-3}}, (b) density profile for multiple supernova injection with SN frequency of $1$ SN in each $3\times 10^4$ year. If we increase the Supernova frequency (i,e one SN in each $1000$ year),then the density profile roughly takes the shape of continuous shell wind-like structure (shown in blue color in the 1st panel of the lower row in figure \ref{fig:supernovae}). The label Central inj. in the figure denotes that CRs are injected at the central region and Shock inj. implies CRs are injected at shocks.}
\label{fig:density}
\end{figure*}

\subsubsection{Cooling}
Radiative cooling that causes thermal energy loss of the gas is non-negligible in dense clouds. To include this, we use
a tabulated cooling function corresponding to collisional equilibrium and solar metallicity (\citealt{Ferland98}, \citealt{Sutherland1993}). 
A floor value in temperature is set to $10^4 K$ so that
cooling is turned off when temperature $T < 10^4$ K, 
which arises from photoionization of the regions in the vicinity of the cluster (\citet{Gupta2016}), on a spatial scale much larger than considered here. The ionized region around Wd1 is larger than the outer radius of the cluster, thereby justifying the floor temperature value of $10^4$ K. If $S_*$ is the number of ionising photons emitted per unit time by the star cluster, $\beta_2$ is the recombination coefficient of hydrogen (Case B approximation), and $n_0$ is the ambient density. The radius of the Str\"{o}mgren sphere be $R_S$ which is given by,
\begin{eqnarray}
    R_S= {\left(\frac{3}{4 \pi} \frac{S_*}{n_0^2 \beta_2} \right) }^{1/3}\,.
    \label{eq:STROM}
\end{eqnarray} 
%
If we use $n_0=50$ cm$^{-3}$, $S_* = 2.26 \times 10^{52}$ s$^{-1}$ (since the ionizing photon luminosity is $\approx 500 L_w$), $T= 10^4$ K, and $\beta_2 = 2 \times 10^{-13}$ cm${^3}$s$^{-1}$, one has 
the radius of Str\"{o}mgren sphere as $\sim 50$ pc. Instead of a uniform medium if we take a $1/r$ -type radial ambient medium then the Str\"{o}mgren radius will be much larger (the forward shock position is only slightly larger) . The net heating is given by
\begin{eqnarray}
{q}_{\rm eff} = -n_i n_e \Lambda_N + {\rm Heating} \,.
\end{eqnarray}
%
The heating of gas due to Coulomb interactions with CRs is negligible because the heating timescale is larger than Gyr. We do not include heating due to CR streaming in our simulations, although we discuss its implications in section \ref{subsec:heating}.

\section{Results}\label{sec:result}
We present our results in this section and then discuss the implications in section \ref{sec:discussion}.


\subsection{Structure of star cluster driven bubble}
Star clusters host massive stars as well as supernova explosions, which produce a low-density bubble around them (\citealt{Weaver77, Gupta2018}). Although the overall size of these bubbles (a few tens of pc) depends mainly on the total mechanical luminosity deposited by the cluster and the ambient density, the interior structure can qualitatively differ depending on whether the energy deposition is dominated by winds or SN explosions (\citealt{Sharma14}). We discuss these differences below.

Figure \ref{fig:density}a shows the density profile of a stellar wind driven bubble (\citealt{Weaver77}) 
at $4.0$ Myr.
There are four distinct regions in the plot: (1) the innermost portion contains the source of energy and mass deposition, (2) the free-wind region where the wind originating from the source expands adiabatically, (3) the shocked-wind region containing slightly more dense gas, (4) the outermost shell containing the swept-up ambient gas. The shocked interstellar medium (ISM) and shocked wind regions are separated by a contact discontinuity (CD). 

Figure \ref{fig:density}b shows the corresponding density profile for multiple supernova injections, where 1 supernova occurs every $3 \times 10^4$ year. For this small rate, we do not observe any wind-like structure in the density profile, but if we increase the supernova frequency, then the density profile does look similar to the case for continuous stellar wind (Blue curve in the 1st panel of the lower row in figure \ref{fig:supernovae}), with four distinct regions as mentioned earlier \citep{Sharma14}.




The size of the bubble, or, to be precise, the distance to the contact discontinuity (CD) is $\approx 80$ pc, for an ambient density of $50 m_H$ cm$^{-3}$. This implies an extended $\gamma$-ray emission region of a similar size. Note that $80$ pc at a distance of $2.8$ kpc subtends an angle of $\approx 98'$. Indeed, the {\it HESS} excess map (\citet{Aharonian2019}, figure 4 in their Supplementary material) shows the $\gamma$-ray bright 
region to have a total extension of $\approx 3^\circ$, consistent with the above estimate for the angular radius. However, we note that roughly half of the last annulus (the fifth) drawn in the same figure by \citet{Aharonian2019} is not $\gamma$-ray bright. This makes the $\gamma$-ray luminosity of the last projected bin comparable to the fourth bin and not brighter, which it would have been, if the $\gamma$-ray bright region had filled the last annulus. At the same time, the morphology of the $\gamma$-ray bright region shows that it is not spherically symmetric. Thus although there is a rough agreement of the size of the bubble (and, consequently, the $\gamma$-ray bright region) from our spherically symmetric simulation with the size of the $\gamma$-ray bright region, a bin-by-bin matching of the simulated result with observations may not be possible.

Indeed, from the structure of the stellar wind bubble (Fig \ref{fig:density}) it is clear that the swept-up shell is much denser than the interior of the bubble. This would result in an enhanced $\gamma$-ray luminosity for the outer radial bin, which would, in turn, dominate the projected luminosity in all projected bins.
\subsection{Different acceleration sites and corresponding observables}
\label{diff_injection}
As mentioned earlier, we consider three different CR acceleration sites 
in our simulations: (1) CR energy injection in the central wind region (using $\epsilon_{\rm cr}$), (2) injection at the shocks (using $w_{\rm cr}$), and (3) combined injection at shocks as well as the central wind region (using both $\epsilon_{\rm cr}$ and $w_{\rm cr}$). We compare our results with the observations of \citet{Aharonian2019}, albeit for a distance of $2.8$ kpc to Wd1 as described in section \ref{sect:distance}. We discuss the effect of 
varying the distance in Appendix A1 .  

\begin{figure*}
\includegraphics[width=\textwidth]{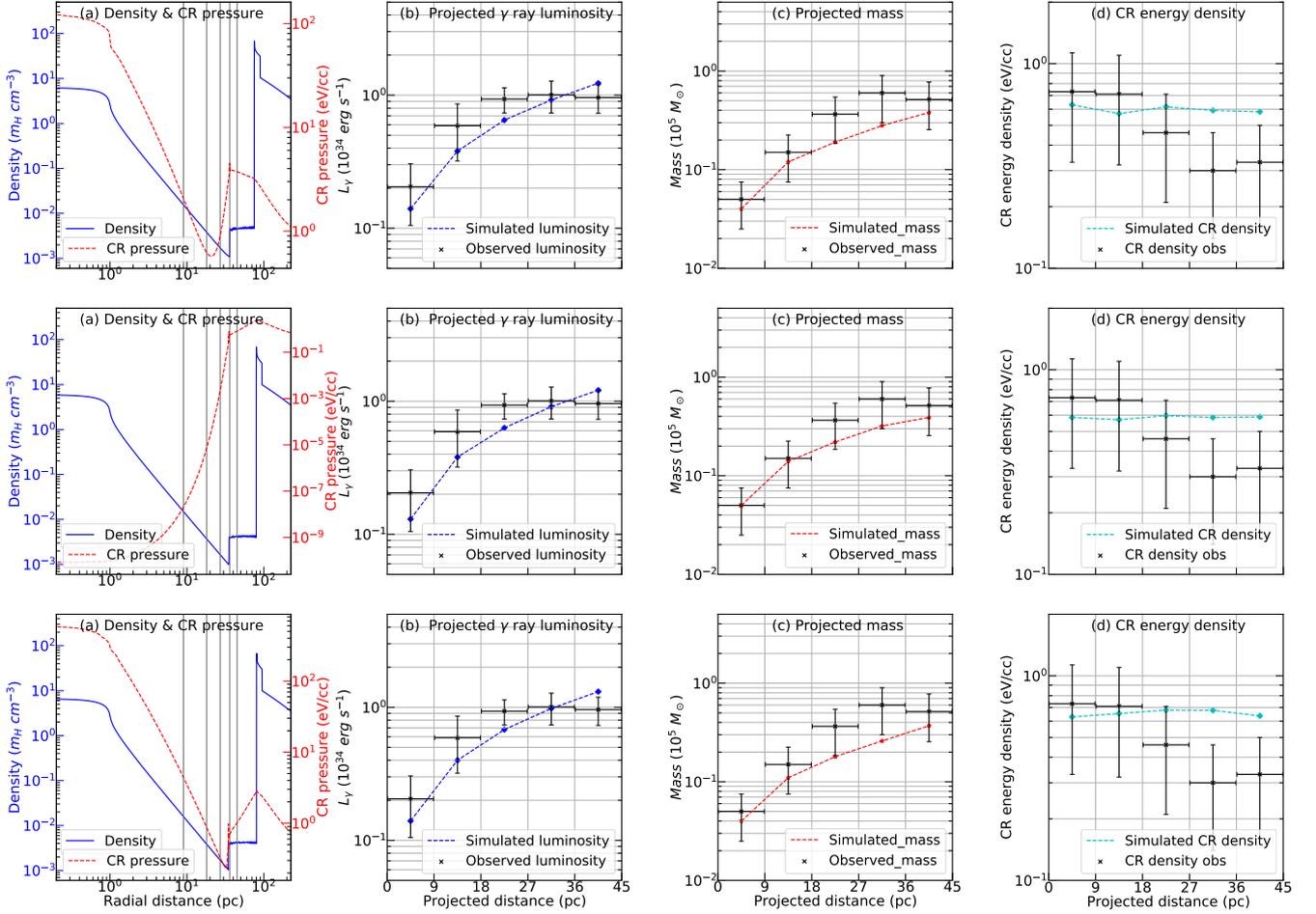}
\caption{
Results of simulations with
the $1/r$ ambient density profile and different injection scenarios are displayed. We plot the
radial density and CR pressure profiles (a), the projected $\gamma$-ray luminosity above 1 TeV (b), projected mass (c), and inferred CR energy density above 10 TeV ($e_{\rm cr,inf}$) (d) for different injection sites of CRs. Black data points with error bars represent observational data and the blue,  red,  and  cyan  dashed  lines  show  the  simulation  results  for  luminosity,  mass,  and  CR  density,  respectively.  The  vertical  lines  in panel (a) represent different projection bins. All profiles are shown at 4.5 Myr. 
The uppermost row shows the case of
central injection with $\kappa_{\rm cr} = 3 \times 10^{27} \si{cm^2 s^{-1}}$ $\epsilon_{\rm cr} =0.1$.  The middle row shows the case of shock injection with 
$\kappa_{\rm cr} = 10^{27} \si{cm^2 s^{-1}}$, $w_{\rm cr} =0.2$. The
bottom row  shows the case of combined injection of CRs, and  for 
$\kappa_{\rm cr} = 10^{27} \si{cm^2 s^{-1}}$, $\epsilon_{\rm cr}=w_{\rm cr} =0.2$. The parameters are chosen to match the $\gamma$-ray luminosity and mass profiles in different scenarios.}

\label{fig:radial_ambient}
\end{figure*}


\subsubsection{Central injection}
In this scenario, CRs are injected into the source region, after which they diffuse outwards. The kinetic luminosity of the stellar wind is distributed in CRs and thermal energy. We define the injection parameter $\epsilon_{\rm cr}$ as,
\begin{eqnarray}
 \epsilon_{\rm cr}=\frac{E_{\rm cr}}{E_{\rm IN}} \,,
 \label{eq:epsilon}
\end{eqnarray}
where $E_{\rm cr}$ is the energy deposited in CRs and $E_{\rm IN}$ is the total deposited energy into the injection region.


We calculated the projected $\gamma$-ray luminosity and mass by dividing the cluster region into $5$ bins from $0-45$ pc with a width of $9$ pc for each bin to compare with the observations of \citet{Aharonian2019}. While calculating the projected $\gamma$-ray luminosity, 
we have 
considered only the hadronic contribution since the leptonic contribution is relatively lower in magnitude than the hadronic contribution.

The fourth column plots the {\it inferred} CR energy density profile\footnote{this is the inferred CR energy density at a given projected distance (details in section 2.2).} in the same manner that observers would have done, based on the projected luminosity and mass. This has been done in order to bring out the essential difference between the actual radial profile (plotted in the first column of Figure \ref{fig:radial_ambient}) and the inferred projected profile of CR energy density, the demonstration of which is the crux of the present paper.

Note that \cite{Aharonian2019} calculated the errors in CR density without considering the uncertainty in the mass estimates (which was mentioned as $\sim 50\%$). 
This has resulted in the underestimation of the errors in the inferred CR energy density. We have, therefore, considered the error in the mass estimates while calculating the final errors in CR energy density. It is found that the revised error bars accommodate a flatter CR energy density profile than expected from a projection of $1/r$ profile. We note that the error mainly arises from the uncertainty in the conversion factor between CO and H$_2$, and its value at length scales as small as $\sim 50$ pc remains unknown.

After exploring the parameter space, we have found that for a $1/r$ type
radial profile of ambient density with the core density of $625$ \si{m_H \,cm^{-3}} as discussed in section \ref{ambient},
the best fit parameters are $\kappa_{\rm cr}= 3 \times 10^{27}$ cm$^2$ s$^{-1}$ and $\epsilon_{\rm cr}=0.1$ (upper row of figure \ref{fig:radial_ambient}).
 
The projected $\gamma$-ray luminosity (above 1 TeV) and mass profiles are shown in the 2nd and 3rd column, respectively in figure \ref{fig:radial_ambient}, for all three different CR injection sites. 
It is clear from this figure that with proper choice of parameters, one can explain the observed values with a $1/r$ -type ambient profile.
Note that we calculated the projected profiles for the whole simulation box i.e. $300$ pc. If we use a simulation box of $400$ pc instead of $300$ pc, the $\gamma$ -ray luminosity changes by $(5-7)\%$.  

\subsubsection{Injection at the shock}
Next, we consider the case of CR injection at strong shocks. We mainly consider injection at the wind termination shock (hereafter WTS) as the Mach number of WTS is much larger than the forward shock (hereafter FS); i.e., WTS is stronger than FS. The efficacy of CR injection at the shocks is described by a commonly used parameter  (\citealt{chevalier83}; \citealt{Bell14}) 
\begin{equation}
    w_{\rm cr} = \frac{p_{\rm cr}}{p_{\rm th}+p_{\rm cr}} \,,
\end{equation}
where $p_{\rm cr}$ and $p_{\rm th}$ are the CR and thermal pressures, respectively (similar as mentioned in section 3). The downstream CR pressure fraction is, therefore, $p_{\rm cr} =w_{\rm cr}\, p_{\rm tot}$ (here, $p_{\rm tot}=p_{\rm cr}+p_{\rm th}$).

After a detailed study of the parameter space, we found that the best fit parameters that can explain the observational data are $\kappa_{\rm cr}=10^{27}$ \si{cm^2 \, s^{-1}}, and $w_{\rm cr}= 0.2$ (consistent with ion acceleration efficiency found in kinetic simulations; see e.g., \citealt{Capiroli14}). 
The middle row of figure \ref{fig:radial_ambient} shows the projected mass and $\gamma$-ray luminosity for these parameters. If we compare with the central injection case (uppermost panel), it is clear that shock injection requires a lower value of $\kappa_{\rm cr}$ than central injection in order to explain the observed  $\gamma$-ray luminosity.


\subsubsection{Combined injection}
We also considered a CR injection scenario where CRs are accelerated in the source region as well as at the shocks. In this case of combined injection, $\epsilon_{\rm cr}$ parametrizes the fraction of kinetic energy that goes into CRs and $w_{\rm cr}$ decides how much of the downstream pressure is converted into CR pressure (same as in section 4.2.1 \& 4.2.2 respectively). The best-matched profiles with observations are shown in the bottom row of the figure 
\ref{fig:radial_ambient}. 
The corresponding value of parameters are $\kappa_{\rm cr}=10^{27}$ \si{cm^2 \, s^{-1}}, and $w_{\rm cr}=\epsilon_{\rm cr}= 0.2$. In table 2, we have mentioned the best fit values of parameters which can explain the observed $\gamma$-ray and mass profile.

\begin{table}
\begin{center}
\renewcommand{\arraystretch}{1.7}
\begin{tabular}{ | c | c| c | } 
\hline
\multirow{2}{4em}{\textbf{Injection sites}} & \textbf{Diffusion co-eff} & \textbf{Inj. parameter} \\ 
 & ($\kappa_{\rm cr}$) \, \si{cm^2 \, s^{-1}} & ($\epsilon_{\rm cr}$ or $w_{\rm cr}$)\\
\hline \hline

\textbf{central injection} & $3 \times 10^{27}$ & 0.1\\ 
\hline
\textbf{shock injection} & $10^{27}$ & 0.2 \\ 
\hline
\textbf{combined injection} & $10^{27}$ & 0.2 \\ 
\hline
\end{tabular}
\end{center}
\caption{\label{tab:table 2}Best fit parameters for different CR injection method.} 
\label{table:best_parameters}
\end{table}

\begin{figure*}
\includegraphics[width=\textwidth]{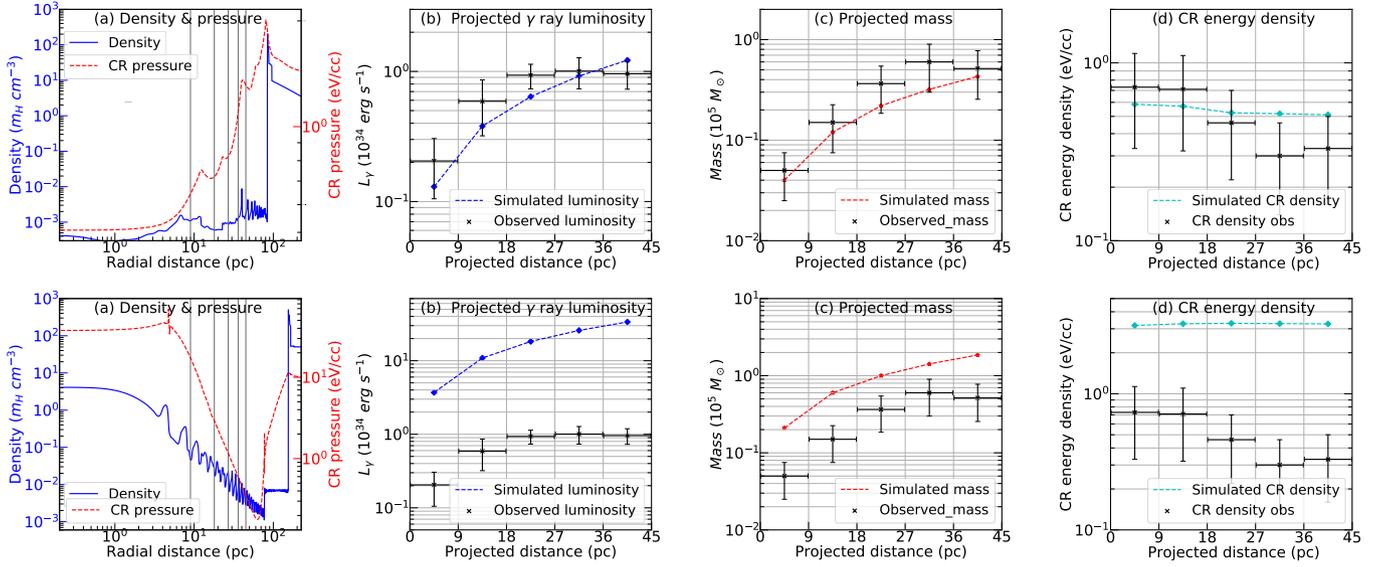}
\caption{Profiles of the density and CR pressure, 
projected $\gamma$-ray luminosity ($>1$ TeV), mass, CR energy density above $10$ TeV for the multiple discrete supernova injection scenario. CRs are injected at the shocks detected by our shock detection method. The value of $\kappa_{\rm cr} = 5 \times 10^{27}$ \si{cm^2\, s^{-1}} and $\epsilon_{\rm cr}=0.1$. Upper row: $1$ supernova in every $0.03$ Myr, lower row: $1$ supernova in every $1000$ year. Only the small supernova rate, consistent with the cluster mass, can satisfy the observational constraints. For the lower panel, we have used a uniform density of $50$ \si{m_H \, cm^{-3}} otherwise, for a $1/r$ type ambient, the forward shock position will be at a very large distance which does not match with the observation.} 
\label{fig:supernovae}
\end{figure*}

\subsection{Multiple discrete supernova injection}
Multiple discrete supernovae (SNe) can also produce stellar wind-like structures if the frequency of supernovae is large (e.g., see Fig. 12 in \citealt{Yadav2017}), and we have considered this alternative as well. For this, the mechanical luminosity $L_w$ will correspond to a kinetic energy of $10^{51}$ erg per SNe, multiplied by the frequency of SNe. \citet{Aharonian2019} suggested a supernova rate of $1$ SN every $1000$ year to support the quasi-continuous injection of CRs in the source region and to explain the observed CR density profile. However, this large rate of SNe is not realistic, because this implies $\approx 3 \times 10^4$ SNe in $30$ Myr (corresponding to the main sequence lifetime of a $8$ M$_\odot$ star), which would correspond to a total stellar mass of $\ge 3 \times 10^6$ M$_\odot$. Therefore, we performed simulations with a more realistic supernova injection frequency of $0.03$ Myr$^{-1}$, 
corresponding to the observed cluster stellar mass of $10^5$ M$_\odot$.  

Figure \ref{fig:supernovae} shows the corresponding projected luminosity, mass and inferred CR energy density profile for multiple supernovae. 
For the above mentioned realistic SNe rate, the density profile, shown in the first panel of upper row of figure \ref{fig:supernovae}, does not show a stellar wind like structure (first panel of the lower row of figure \ref{fig:supernovae}) , which is achieved only for a high rate of supernova ({\it e.g.,} $1$ SN in every $1000$ yr) (lower row of fig \ref{fig:supernovae}). Yet, one can get a close enough match with the projected luminosity and mass profiles. The best fit parameters for SNe rate of $0.03$ Myr$^{-1}$ are $\kappa_{\rm cr} = 5 \times 10^{27}$ \si{cm^2 \, s^{-1}}, $\epsilon_{\rm cr}=0.1$, for an $1/r$ -type ambient density.  When the SNe rate is increased, the corresponding luminosity, mass, and inferred CR energy density profile much exceed the observed values. We have included this high rate of supernova just to look at the prediction of \citet{Aharonian2019} assumption. Also, for this high SNe rate, we have used a uniform ambient medium of $50$ \si{m_H \, cm^{-3}}. Instead, if we use a 1/r type ambient medium, the outer shock position will be at a large distance (beyond $\sim 220$ pc) which does not match the observation. 



\begin{figure}
\includegraphics[width=80mm]{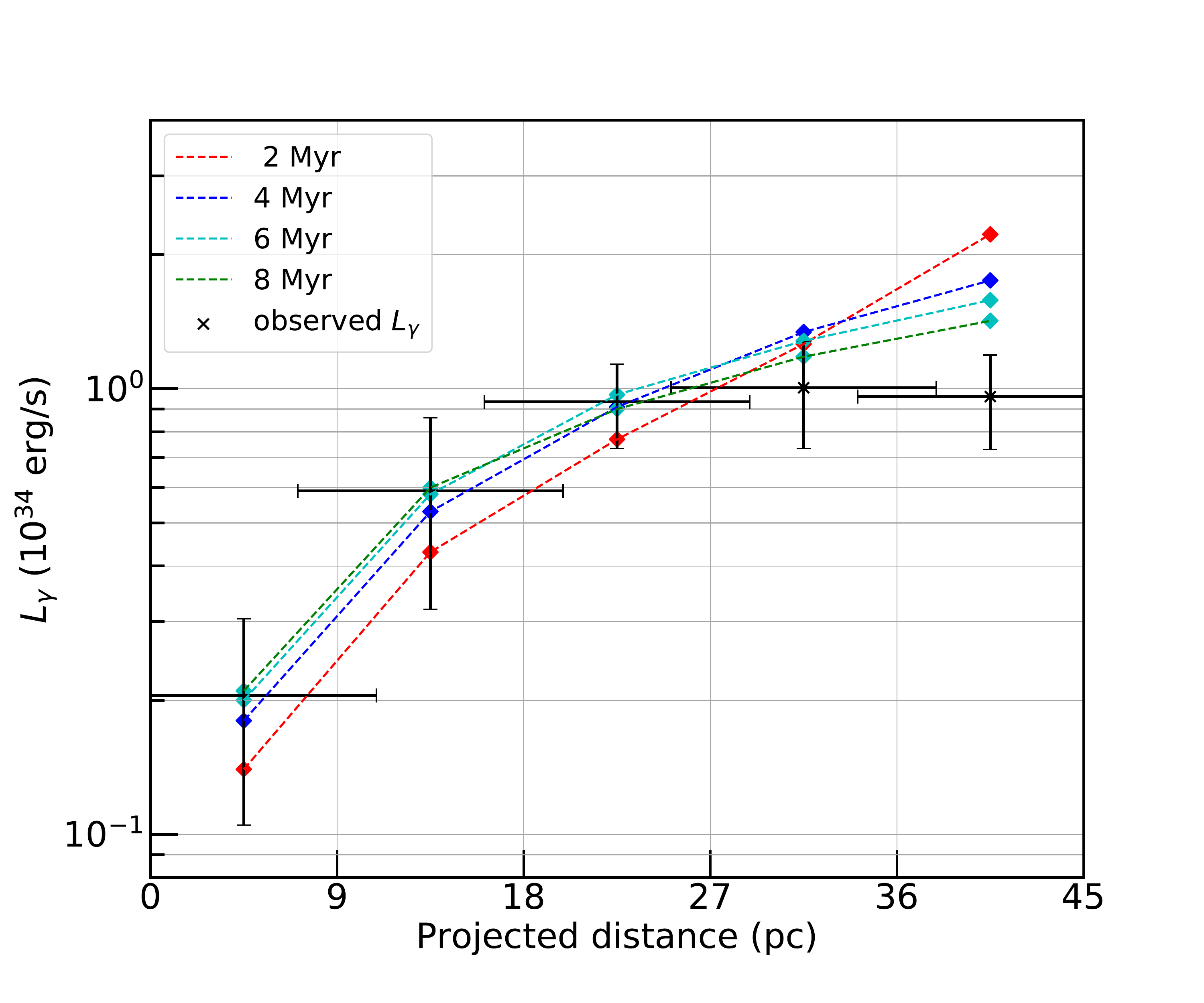}
\caption{Time evolution of $\gamma$-ray luminosity for combined injection with $\kappa_{\rm cr}=10^{27}$ \si{cm^2 \, s^{-1}}, $w_{\rm cr}=\epsilon_{\rm cr}=0.2$. Black points are from observation.} 
\label{fig:time_evolution}
\end{figure}

\section{Discussion}
\label{sec:discussion}
Our simulated $\gamma$-ray luminosity and mass profiles match the observations (panels (b) and (c) of figure \ref{fig:radial_ambient}), for the parameter values mentioned in each case of CR injection. We also note that the inferred CR energy density offers a good match with the observed profiles, in light of the revised error bars that include the uncertainty in mass estimation (panel (d) in each row of figure \ref{fig:radial_ambient}).  
It should also be noted that our simulations are based on some simple assumptions e.g., spherical symmetry and constant diffusion coefficient.
3-D simulations can produce more realistic morphology, but those require additional free parameters such as mass distribution of cloud and location of stars.
It is, therefore, reasonable to say that the present simulations offer a good match with the observations, in light of all the uncertainties mentioned earlier.

There are other circumstantial reasons why a flatter CR energy density profile should be considered. Recently \citet{Abeysekara2021} have shown (in their figure 2b) that for Cygnus cluster, the CR energy density above $10$ \si{TeV} does not strictly follow a $1/r$ profile, and their observation does not rule out $e_{\rm cr,inf}$ being uniform, 
which would make it consistent with our simulation results (panel (d) in each row of figure \ref{fig:radial_ambient}). 
At the same time, the CR energy density profile for $100$ \si{GeV} does follow $1/r$ profile \citep{Aharonian2019}. \citet{Abeysekara2021} interpreted this absence of a $1/r$ profile for \si{TeV} CRs on the basis of larger diffusion rate for higher energy CRs. 

The comparison of $L_\gamma$ and $e_{\rm cr,inf}$ from our simulation and observations indicate that the last projected bin is observed to be less luminous than expected from simulation. There can be a variety of reasons for this discrepancy. One possibility is that the outer shell is fragmented and is porous, as in the case of 30 Doradus, for example (which allows the X-ray from the shocked wind region to be seen through the holes in the outer shell). Such a fragmented outer shell may make the $\gamma$-ray luminosity in the outer-most bin discrepant from the simulated values.

\subsection{Time dependence of gamma ray profiles}
Figure \ref{fig:time_evolution} shows the time dependency of the $\gamma$-ray luminosity profile for combined injection of CRs. As time increases, the bubble structure expands. The luminosity in the inner bins increases with time, but the outer bin shows an opposite trend. This is because, as time progresses, the outer shock covers a more extended region, thereby increasing the effective volume of the emitting region, and increasing the luminosity in the inner bins, because of the projection effect. At the same time, the WTS and the shocked wind region gradually move out of the intermediate and outer bins, thereby decreasing the contributions in luminosity in those bins. The difference in the luminosity from $2$ to $4$ Myr is found to be roughly $\sim 25\%$, and within the observational margin of error. 

\begin{figure}
\includegraphics[width=3.5in]{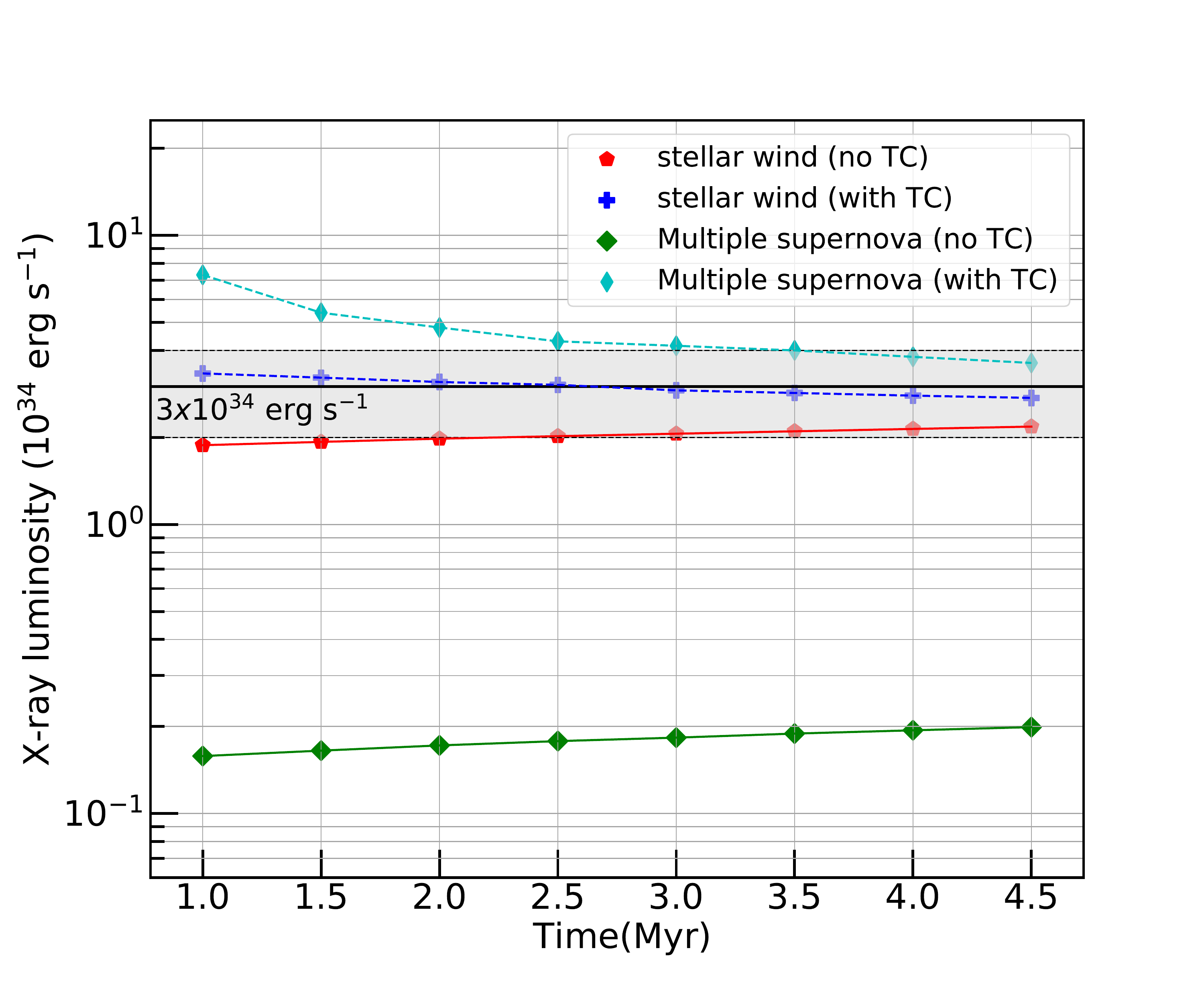}
\caption{Time evolution of X-ray luminosity ($2-8$ keV) for stellar wind case (red and blue curves) and multiple supernova case (green and cyan lines). The black solid line shows 
$3 \times 10^{34}$ erg/s 
which is the obtained value from observation. The shaded region shows the range of the observed luminosity. 
Solid and dashed curves correspond to without TC runs and with TC runs respectively.} 
\label{fig:xray}
\end{figure}

\subsection{Effect of thermal conduction}
We have also studied the effect of thermal conduction on the simulated $\gamma$-ray profiles. We use thermal conduction to have the spitzer value ($\kappa_{\rm th} = 6 \times 10^{-7} T^{5/2}$ in CGS, \citealt{Spitzer1962}) and also assumes the saturated thermal conduction (section 4.3 of \citealt{Gupta2016}). For the two-fluid model, thermal conduction does not significantly change the simulated $\gamma$-ray profile, and the change in the $\gamma$-ray luminosity in each bin is $\le (5-7)\%$. 

\subsection{Thermal X-rays}

We have calculated the resulting X-ray luminosity of the (hot and dense) shocked wind region. We consider the X-ray emission due to thermal bremsstrahlung, which can be calculated using (equation 5.14b of \citealt{rybicki}),
\begin{eqnarray}
    L_x = \int_V dV \, \int_\nu d\nu\left[ 6.8 \times 10^{-38} \, Z^2 \, n_e \, n_i T^{-1/2}\, e^{-h\nu/kT}\tilde{g_{\rm ff}}\right]
\end{eqnarray}

We take $n_e \sim n_p = P_{\rm th}/k_B T$ ($P_{\rm th}$ is the thermal pressure), $Z \sim 1$, and $\tilde{g_{\rm ff}}=1.2$. The X-ray luminosity in $2-8$ keV for both stellar wind and multiple supernova cases is shown in figure \ref{fig:xray}. For stellar wind scenario the X-ray luminosity matches the observed value \citep{Muno2006} of $(3\pm 1) \times 10^{34} \si{erg \, s^{-1}}$ (shown by the shaded region) with or without thermal conduction. However, the corresponding X-ray luminosity for the multiple supernova injection scenario is more than one order lower in magnitude than the observed value if we do not include thermal conduction. This 
is due to the very low density of the gas (see figure:\ref{fig:density}b) inside the bubble, owing to the low SNe rate. (Higher SNe rate would recover the density structure, but overproduce $\gamma$-rays, as shown in the bottom panel of figure \ref{fig:supernovae}.) 
However, with thermal conduction, the simulated values (cyan curve in the plot) are close to the shaded region for this injection scenario. For this case, we have set an upper limit of conduction coefficient, which corresponds to $10^7$K temperature (otherwise, the stability timescale due to thermal conduction is too short). Therefore, both the stellar wind and multiple supernova injection models can explain the observed $\gamma$-ray as well as X-ray luminosity. 


\subsection{Heating due to CRs and CR energy loss}
\label{subsec:heating}

CR energy loss, due to Coulomb and hadronic interactions, can indeed be important, as estimated below. Using the expressions for Coulomb and hadronic loss in \citet{Guo2008}, the total CR energy loss rate is,
\begin{eqnarray}
    \Gamma_c = 7.6\times 10^{-16} \left (\frac{n}{\si{cm^{-3}}} \right) \, \left (\frac{e_{\rm cr}}{\si{erg \, cm^{-3}}} \right) \, \si{erg \, s^{-1} cm^{-3}} \, .
\end{eqnarray}
The heating time for the gas is,
\begin{eqnarray}
    t_H \approx \frac{1.5 nkT}{(1.65 \times 10^{-16} \, n \, e_{\rm cr})} \, \si{sec}\,,
\end{eqnarray}
considering only the Coulomb interaction. Using $e_{\rm cr} \sim $ 0.45 \si{eV cm^{-3}},  $T=10^4$ K (corresponds to shell temperature),  the heating time scale is $t_H \sim 10^{9}$ \si{yr}. This heating time is much larger than the dynamical time scale of $4.5$ Myr, so the effect of this heating is negligible for the thermal gas. 
However, the energy loss time scale for CRs is,
\begin{equation}
    t_{\rm cr, loss}\approx 0.4 \, {\rm Myr} \, \Bigl ({ n \over 50 \, \rm m_{H} \, {\rm cm}^{-3}} \Bigr )^{-1} \,.
    \label{cr-loss}
\end{equation}
We can also estimate the energy loss due to CR streaming heating, for which the heating rate is given by,
\begin{eqnarray}
    \Gamma_{\rm streaming} = - v_A \cdot \nabla p_{\rm cr} \, \, \si{erg \, cm^{-3}s^{-1}} \, .
\end{eqnarray}
Here, $p_{\rm cr}$ is the CR pressure and $v_A$ is the Alfven velocity. 
If we assume equipartition of magnetic and thermal energy density, then 
$v_A \approx 1.3\times 10^8$ \si{cm \, s^{-1}}. 
If we consider the region between $20\hbox{--}50$ pc in the density plot (panel (a) of the topmost row of figure \ref{fig:radial_ambient}, we  find that the change of CR pressure ($\Delta p_{\rm cr}$) is $\approx 1.8 \times 10^{-11}$ \si{dyne \,cm^{-2}} over a  distance ($\Delta r$) of $30$ pc. 
This gives us, $\Gamma_{\rm streaming} \approx  2.6 \times 10^{-23} \, \si{erg \, cm^{-3}s^{-1}}$. The energy loss timescale for CR is long, but the heating time scale for the gas is  $\sim 0.2$ \si{Myr}, (for $n\approx 0.01$ \si{cm^{-3}}). Although this may be important, we have not included streaming heating in our simulations because it will involve making assumptions about the uncertain small-scale magnetic fields.

The above discussion, especially regarding the energy loss time scale for CR (equation \ref{cr-loss}), shows that CR energy density in the shocked wind and outer shell can significantly decrease over the considered dynamical time scale. This process would reduce the CR energy density in these regions and consequently decrease $L_\gamma$. Therefore, $L_\gamma$ would be lower than presented here, especially in the outer bins, and make the inferred CR energy density decline with the projected distance. This may result in a better match with the observations.

Our analysis shows that the diffusion coefficient  ($\kappa_{\rm cr}$) lies  in the range of $(5-30) \times 10^{26}$ \si{cm^2 \, s^{-1}}. Note that CR diffusion is ineffective for a much lower diffusion
coefficient, whereas CRs rapidly diffuse out of the bubble without affecting it if $\kappa_{\rm cr}$ is
increased (see also \citet{Gupta2018}). A comparison of the simulation results with observation implies that the $\gamma$ -ray luminosity matches well if CR energy fraction $ 10-20 \%$  of the total input energy, consistent with theoretical expectations from diffuse shock acceleration mechanisms.

\begin{figure*}[ht!]
\includegraphics[width=\textwidth]{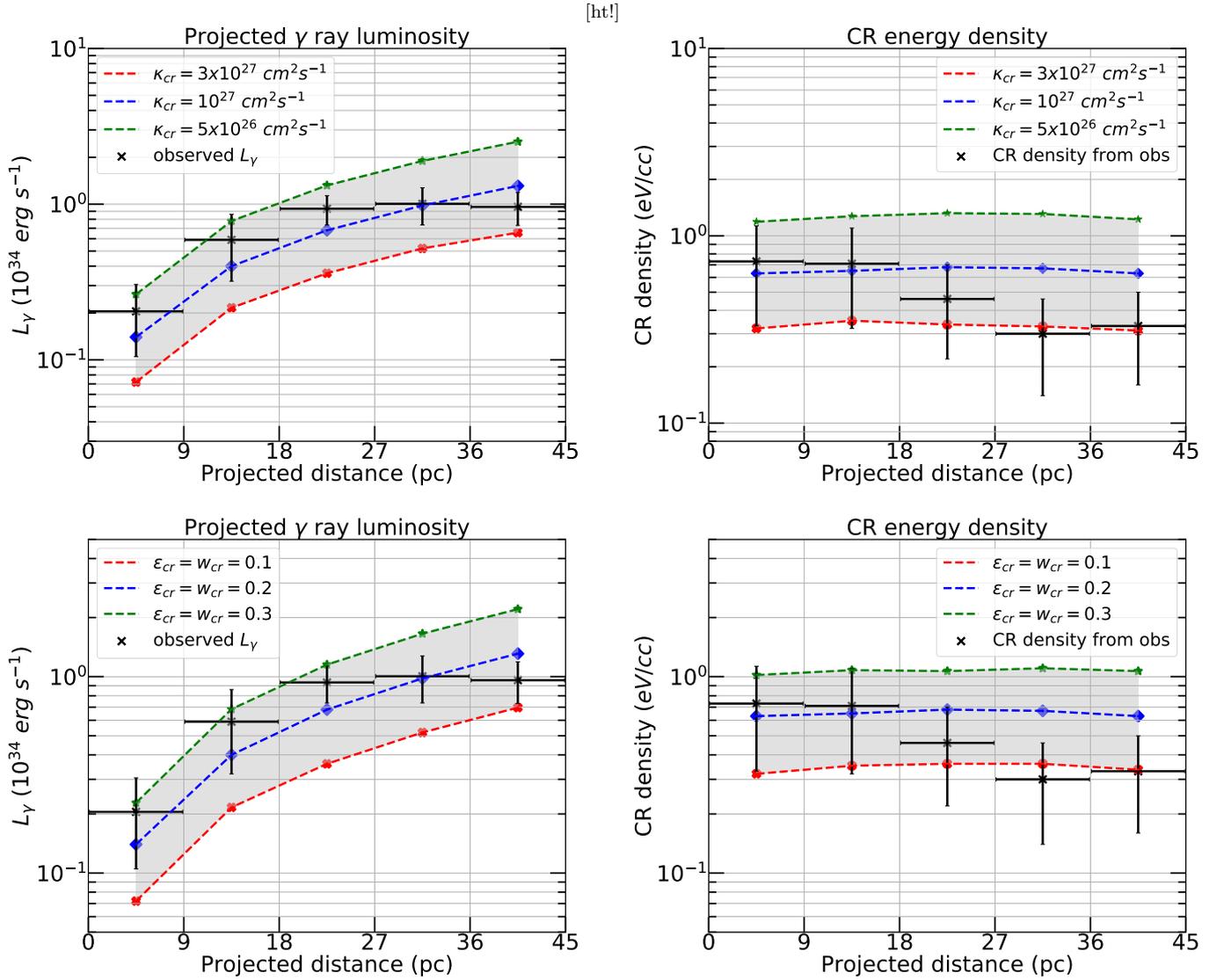}
\caption{The projected $\gamma$-ray luminosity and inferred CR energy density profiles as a function of the projected radius for different injection parameters and diffusion parameters for the case of combined CR injection scenario. In all panels, black points with error bars indicate the observational values. The upper left and right panels show the variation of the 
$\gamma$-ray luminosity and inferred CR energy density for different $\kappa_{\rm cr}$, respectively for a fixed $w_{\rm cr}= \epsilon_{\rm cr}=0.2$.
The lower left panel shows the variation of the 
$\gamma$-ray profile with varying CR injection parameter and the lower right panel shows the variation of the 
projected inferred CR energy density profile with varying injection parameter. For the lower two panels the value of $\kappa_{\rm cr}= 10^{27}$ \si{cm^2 s^{-1}}.}
\label{fig:comparisons}
\end{figure*}

Also, note that the $\gamma$-ray luminosity is a function of both gas density ($n_N$) and CR energy density ($e_{\rm cr}$), whereas the mass is only a function of gas density. As $e_{\rm cr}$ depends on $\kappa_{\rm cr}$,  the $\gamma$-ray luminosity changes significantly with a change in the diffusion coefficient, as shown in the upper left panel of figure \ref{fig:comparisons}. In contrast, the mass profile does not strongly depend on our choice of parameters. For example, although the size of a stellar wind bubble depends on $\epsilon_{\rm cr}$, the projected mass does not change noticeably like the $\gamma$-ray luminosity for different values of $\epsilon_{\rm cr}$.

\subsection{Dependence on various parameters}
We have studied the dependence of our results on different parameters, {\it viz.}, the diffusion coefficient and the injection parameters.

\subsubsection{Diffusion coefficient ($\kappa_{\rm cr}$)}

To understand the effect of diffusion coefficient on the $\gamma$-ray profile, we also ran the simulations for different values of the diffusion coefficient, keeping a constant $\epsilon_{\rm cr}= w_{\rm cr}=0.2$. The upper left and right panel of figure \ref{fig:comparisons} respectively show the variation of $\gamma$-ray luminosity and CR density with distance for different values of $\kappa_{\rm cr}$. 
It is clear from the upper left panel of the same figure, the $\gamma$-ray luminosity exceeds the observed values for a lower value of diffusion coefficient. This is because a slower diffusion of CRs implies a higher density of CRs in the vicinity of the cluster, which increases the $\gamma$-ray luminosity.

The upper left panel of figure \ref{fig:comparisons} shows the corresponding variation of the inferred CR energy density profile with $\kappa_{\rm cr}$. As expected, increasing the diffusion coefficient depletes the injection region of CRs, and the resulting drained CR energy density profile is naturally decreased. However, our exercise selects the range of $\kappa_{\rm cr}\approx (5\hbox{--}100) \times 10^{26}$ cm$^2$ s$^{-1}$ as the appropriate one since the observed values are bracketed from both sides in this range, as seen from the 
the upper left and upper right panels of figure \ref{fig:comparisons}. We note that this range of $\kappa_{\rm cr}$ is consistent with previous estimates from observations of $\gamma$-rays in star clusters (\citealt{Gabici2010}; \citealt{Giu2010}; \citealt{Li2010}; \citealt{Ackermann2011}).

\subsubsection{Injection parameter ($w_{\rm cr}$ \&  $\epsilon_{\rm cr}$)}
We have also run the simulations for different values of the injection parameter ($w_{\rm cr}$ and $\kappa_{\rm cr}$) keeping a constant diffusion coefficient $\kappa_{\rm cr}=  10^{27}$ cm$^2$ s$^{-1}$. The lower-left and right panel of figure \ref{fig:comparisons} respectively shows the variation of $\gamma$-ray luminosity and CR density with distance for different injection parameters. It is clear from the lower-left panel of the figure that an increasing value of $w_{\rm cr}$ or $\epsilon_{\rm cr}$ increases the $\gamma$-ray luminosity, because a larger injection parameter means a larger fraction of kinetic energy being deposited into CRs which consequently increases the $\gamma$-ray luminosity in the close vicinity of the cluster. The corresponding CR density profile is shown in the lower right panel of the figure \ref{fig:comparisons}. 

\section{Conclusions}
We have studied the implication of the recently inferred distribution of CR energy density in massive compact star clusters, taking the particular example of Wd1.
With 1-D two-fluid hydro-dynamical simulation for stellar wind in star clusters, we have studied the projected $\gamma$-ray luminosity, mass and CR energy density for the Wd1 cluster and their dependence on diffusion coefficient, injection parameter and ambient density. 
Our findings are as follows: 

\begin{enumerate}
\item The most important takeaway from our analysis is that the inferred $1/r$ profile of CR energy
density need not reflect its true radial profile. Also, we have shown that even the observed data can accommodate a flatter CR energy density profile, in light of revised error estimates. We have shown that dividing the projected $L_\gamma$ by the projected mass in different annuli can yield a CR energy density profile that is significantly different from the actual profile. We have also pointed out various uncertainties that would make a straightforward inference difficult, {\it e.g.}, the lack of morphological symmetry, the uncertainty in the mass estimate. 

\item  While a $1/r$ profile for the CR energy density allows a simple explanation in terms of a steady-state CR luminosity at the centre of the cluster, which makes it appealing, we have studied the 
more plausible scenarios, that of a time-varying CR luminosity, or CR being injected outside the central region (in the wind termination shock, for example), and showed how these scenarios 
are also consistent with observations.
We can not rule out any of the CR acceleration 
sites on the basis of these observations because the observed luminosity and mass profile can be explained by all three CR injection methods, as well as the discrete supernova scenario by appropriate choice of the diffusion coefficient and injection parameters. 
\item The parameters for the best match with observations are not ad-hoc, but are supported by independent arguments. For example, a lower value of diffusion coefficient ($ 10^{27}$ cm$^2$ s$^{-1}$) can explain the observation for shock injection case, while for central injection a higher value ($3 \times 10^{27}$ cm$^2$ s$^{-1}$) is required. These values for the diffusion coefficient are consistent with previous findings. 
The same goes for the parameter describing the efficiency of CR energy injection, which is found to be in the range $\epsilon_{\rm cr}$/$w_{\rm cr}$ $\sim 0.1\hbox{--}0.3$, consistent with previous works \citep{Gupta2018}. 
\item The discrete multiple supernova injection scenario can explain the $\gamma$-ray observation with the appropriate choice of parameters. On the other hand, the simulated X-ray luminosity (assuming it to be thermal) is close to the observed value only if we include thermal conduction. 
\end{enumerate}

\section{Data availability}
The data not explicitly presented in the paper will be available upon reasonable request from the first author.

\section{Acknowledgements}
We would like to thank Manami Roy, Ranita Jana, Alankar Dutta for the valuable discussion. BBN would like to thank Felix Aharonian, Ruizhi Yang, Binita Hona for useful discussion on the data. We thank our anonymous referee for helpful comments. SB acknowledges Prime Minister's Research Fellowship (PMRF), Govt. of India for financial support. PS acknowledges a Swarnajayanti Fellowship (DST/SJF/PSA-03/2016-17) and a National Supercomputing Mission (NSM) grant from the Department of Science and Technology, India.


\section*{Appendix A1: Effect of distance of WD1}\label{appendix:A1}

In the present work, we have used a distance of $2.8$ kpc for Wd1, but it is important to know how the difference in distance affects the inferences, because previous works assumed it to be $4$ kpc. For this reason, we have determined the projected $\gamma$-ray luminosity and mass for a distance of $4$ kpc (which is $1.4$ times larger than the new predicted distance of $2.8$ kpc). As a consequence of this, the observed $\gamma$-ray luminosity, as well as the projected mass, will increase by a factor of $1.4^2 \approx 2$ for each bin. Also, the width of the bin will increase $1.4$ times and each bin width will become $13$ pc instead of $9$ pc that we have used in our calculations in the main text. If we consider a bin of projected distance between $w_1$ and $w_2$, then the total projected luminosity in the bin is calculated by integrating over this region (i.e from $w_1$ to $w_2$),
\begin{figure*}
\renewcommand{\thefigure}{A1}
\includegraphics[width=\textwidth,height=5cm]{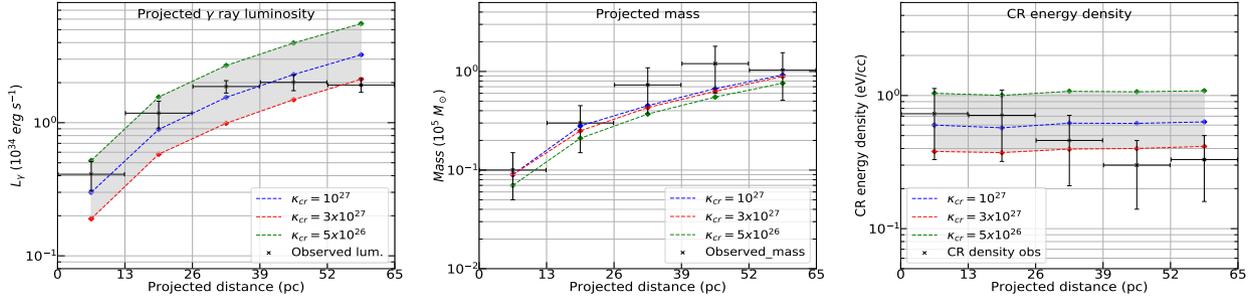}
\caption{Same as in the bottom row of Figure \ref{fig:radial_ambient} except that a distance of WD1 is taken to be $4$ kpc. The effect of varying $\kappa_{\rm cr}$ is also shown in the figure.} 
\label{4kpc distance}
\end{figure*}

\begin{eqnarray*}
    L_{w_1\rightarrow w_2} &=& 2 \pi \int_{w_1}^{w_2} \left[{\int_{w_1}^{r_{box}}2 j_{\nu}(r)  \frac{r dr}{\sqrt{r^2-w^2}} \hspace{3pt}}\right] w \hspace{3pt} dw \, ,
\end{eqnarray*}

where $j_{\nu}$ is the emissivity, $r_{\rm box}$ is the maximum box size and $r$ is the radial distance.
A change in the distance will modify the bin width and will affect the $\gamma$-ray luminosity through the above integral. Since the projected mass is calculated in a similar way, the mass estimate will also change in a similar way.

Figure \ref{4kpc distance} shows the corresponding profiles for a distance of $4$ kpc distance, for the case of combined CR injection. Upon comparison with the bottom panel of figure \ref{fig:radial_ambient}, we find that the modified projected $L_\gamma$ and projected mass are still within the observational error bars. 
The CR energy density does not deviate much from the limit of error bar of the data points. To summarise, the effect of changing the distance is rather modest in light of the observational uncertainties and does not significantly affect the conclusions.

\section*{appendix a2: Effect of time-varying mechanical luminosity: Starburst99}
\begin{figure*}
\renewcommand{\thefigure}{A2}
\includegraphics[width=\textwidth,height=5cm]{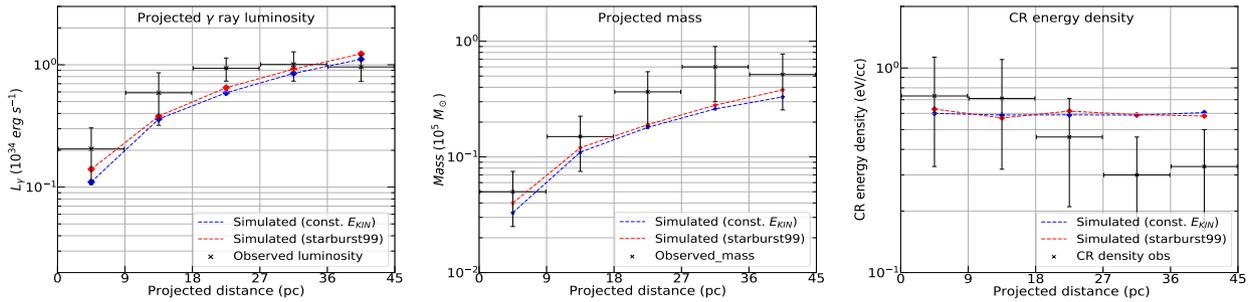}
\caption{Variation of projected $\gamma$-ray luminosity above 1 TeV, mass and CR energy density above 10 TeV as a function of projected distance for central CR injection scenario. The red curve is for Starburst99 model and the blue curve is for constant mechanical luminosity driven model. The match between the two is very close. 
} 
\label{fig:starburst}
\end{figure*}
We have also investigated the effect of the time-dependent mechanical luminosity of the cluster, $L_w$, using Starburst99 
\footnote{\href {https://www.stsci.edu/science/starburst99/docs/default.htm}{https://www.stsci.edu/science/starburst99/docs/default.htm}} \citep{Leitherer99}, which is a publicly available code for stellar evolution in clusters. 
We use the Padova AGB track with solar metallicity and instantaneous star formation for this calculation. 
In figure \ref{fig:starburst} we compare our result with the case of a constant mechanical luminosity-driven wind model.
We find that these two models do not differ much in terms of projected luminosity, mass, or inferred CR energy density. Here the parameters used are $\kappa_{\rm cr}= 3 \times 10^{27}$ \si{cm^2 s^{-1}}, $\epsilon_{\rm cr}= 0.1$ for central injection of CRs. 

\end{document}